\documentclass[
preprint,
bibnotes,
 amsmath,amssymb,
 aps,
 amsthm,
floatfix]{revtex4-1}

\usepackage{graphicx}
\usepackage{dcolumn}
\usepackage{bm}
\newtheorem{thm}{Theorem}
\newtheorem{ass}{Assumption}
\newtheorem{lem}{Lemma}

\begin{document}

\preprint{APS/PRE}

\title{Stability of a planetary climate system with the biosphere competing for resources}

\author{Sergey A. Vakulenko,$^{1,2}$ Ivan Sudakov,$^{3,*}$ Sergei V. Petrovskii,$^4$ and Dmitry Lukichev$^{2}$}
\affiliation{$^1$Institute of Problems in Mechanical Engineering, Russian Academy of Sciences, St.\,Petersburg 199178, Russia\\
$^2$Department of Electrical Engineering and Precision Electro-Mechanical Systems, ITMO University, St.\,Petersburg 197101, Russia \\
$^3$Department of Physics, University of Dayton, 300 College Park, SC 111, Dayton, Ohio 45469-2314, USA\\
$^4$School of Mathematics and Actuarial Science, University of Leicester, Leicester LE1 7RH, UK}
\email{Corresponding author: isudakov1@udayton.edu}

\begin{abstract}
With the growing number of discovered exoplanets, the Gaia concept finds its second wind. The Gaia concept defines that the biosphere of an inhabited planet regulates a planetary climate through feedback loops such that the planet remains habitable. Crunching 'Gaia' puzzle has been a focus of intense empirical research. Much less attention has been paid to the mathematical realization of this concept. In this paper, we consider the stability of a planetary climate system with the dynamic biosphere by linking a conceptual climate model to a generic population dynamics model with random parameters. We first show that the dynamics of the corresponding coupled system possesses multiple timescales and hence falls into the class of slow-fast dynamics. We then investigate the properties of a general dynamical system to which our model belongs and prove that the feedbacks from the biosphere dynamics cannot break the system's stability as long as the biodiversity is sufficiently high. That may explain why the climate is apparently stable over long time intervals. Interestingly, our coupled climate-biosphere system can lose its stability if biodiversity decreases; in this case, the evolution of the biosphere under the effect of random factors can lead to a global climate change.
\end{abstract}


\maketitle


\section{Introduction} \label{intro}
Understanding of the mechanisms and scenarios of climate change as well its current and potential effects on ecosystems and biodiversity have been a focus of keen attention and intense research over the last few decades \cite{Chen14,IPCC,IPCC14}. There is a general consensus that climate change will likely have an adverse impact on the ecological systems and population communities resulting in species extinction and a considerable biodiversity loss worldwide. 

Whilst the top-down effect of climate on ecosystems is thus well established, relatively little attention has been paid to a possibility of an opposite, bottom-up effect that ecosystems may have on the climate. The mainstream of research often tends to consider the ecosystems and population communities as ‘biological actors on the physical stage’ \cite{Mortimer75} often disregarding possible feedback. 
Meanwhile, in planetary science, there is the concept of Gaia \cite{Love79} that postulates the biosphere regulates its planetary climate to mitigate it for its own survival. While this hypothesis has been introduced quite long ago, current research in planetary and earth sciences inspires new applications of this hypothesis. The work \cite{Ab20} is shown that even if a model exoplanet has significant climate perturbations then the Gaia concept is still acceptable (the original Gaia concept is based on a static planetary climate). Another work \cite{Lent18} supports the Gaia concept considering Earth's biosphere stability over climate change through the existence of climate feedback loops and climate tipping points \cite{ Lent, Ash}.  

In this paper, we present a new mathematical realization of the Gaia hypothesis by a model of coupled climate-biosphere dynamics. We consider the effect that the biosphere of a planet may have on a planetary climate by changing the global energy balance through modifying the planetary albedo.
\raggedbottom

Modeling of physical processes in the climate system leads to difficult problems, involving complicated systems of partial differential equations for biological and chemical processes \cite{IPCC}. There exist climate models with different levels of realism; they can include thousands and even millions of equations, thousands of parameters to adjust. Usually, one investigates these models by computer simulations \cite{Hur}. However, it is difficult to estimate the reliability of these computations, since it is connected with a difficult mathematical problem on the structural stability of attractors \cite{Katok, smale}. The theory of linear response of climate systems to perturbations \cite{Luc} is based on the Ruelle theory of linear response for dynamical systems that holds on the formal hypothesis that the dynamical system is of the type axiom $A$ one. The last fact implies structural stability. However, S. Smale's $A$-axiom systems \cite{smale} seldom appear in practical applications. The class of structurally stable systems is very narrow; this mainly includes systems with hyperbolic or almost hyperbolic behavior. One can expect, therefore, that the attractors of climate systems are not structurally stable: their topological structure can change under small perturbations. Therefore, one can expect that they can exhibit complicated bifurcations under small parameter perturbations.  Possibly, an adequate approach is to take into account random fluctuations and study random dynamical systems.
Indeed, the climate system as a complex system has a large number of interconnected and interacting subsystems, including the following: the atmosphere, the oceans, the biosphere, etc. Determination of how the dynamics of these subsystems change to reach equilibria of the entire system is the main problem of so-called conceptual climate models. 

There are different types of conceptual climate models. Many of them are energy balance climate models and are defined by an ordinary differential equation describing energy conservation in the climate system. The most popular model is a {\it zero-dimensional model} \cite{North} based on the theory of blackbody radiation determining global temperature changes due to the difference in incoming and outgoing solar radiation. This difference may be caused by changing of control parameters: surface albedo, greenhouse gas emission, and even the solar constant. The equilibria and the ideas how to find them by the bifurcation theory tools can be found here \cite{Sellers}.


In the context of Gaia hypothesis \cite{Ash, Lent} here arises a key question: Why does climate stays stable over long time intervals (e.g.~hundreds of thousands of years)? To answer this question, we consider conceptual climate models where the dynamical variables may be decomposed as slow and fast modes. Then for large times fast mode  dynamics is captured by the slow dynamics on a stable slow manifold of a slow-fast system. The slow variables determine a long-term climate evolution under external factors whereas the fast modes may be associated with rapid factors.

The paper is organized as follows. In the next section, we introduce a planetary climate model with a biosphere component that arises from coupling between the conceptual zero-dimensional global energy balance model of climate dynamics and a generic ecosystem dynamics model (a multispecific population system living on multiple food sources). In Section \ref{sec:general}, we consider a general class of systems to which our model belongs and discuss the stability of those systems. We then show in Section \ref{sec:oursystem} that, in the case of our climate-biosphere model, a planetary climate remains stable with regard to a variation of the ecosystem model parameters as long as biodiversity is sufficiently large but it can lose stability (hence potentially resulting in regime shifts and a global climate change) if the number of species is small. A discussion and conclusions can be found in the last section.

\section{The model}\label{modex}

The energy balance system is one of simplest climate models. It is defined by the following equation \cite{Sellers} :
\begin{equation} \label{sellers}
\frac{dT}{dt}=\lambda^{-1} \left(-e \sigma T^4 + \frac{\mu_0 I_0}{4}
(1- A)\right),
\end{equation}
where $\lambda$ is thermal inertia, $T$ is the averaged surface temperature, $t$ is time, and $A$ is the albedo of the surface. The left term characterizes the time-dependent behavior of the climate system.  On the right hand side, the first term is the outgoing emission and the second term represents the incoming star's radiation. Generally, incoming radiation to the planetary surface from a star is modified by a parameter, $\mu_0$, to allow for variations in the stellar irradiance per unit area, $I_0$ (the solar constant in case of the Earth), or for long-term variations of the planetary orbit \cite{astr}. On the other side, the outgoing emission depends on the fourth power of temperature, the effective emissivity $e$ and a Stefan-–Boltzmann constant $\sigma$.

This model can be coupled with the modeled biosphere's dynamics as follows. The complete averaged albedo $A$ can depend on the biosphere state. For simplicity, we mostly focus our analysis on a single global ecosystem competing for several resources. 
We consider the following classical model:
\begin{equation}
     \frac{dx_i}{dt}=x_i (- \mu_i  + \phi_i(v) -  \gamma_{i} \; x_i), \quad i=1,\dots, m,
    \label{HX1}
     \end{equation}
\begin{equation}
     \frac{dv_k}{dt}=D_k(S_k -v_k)   -  \sum_{i=1}^M b_{ ki} \; x_i \; \phi_i(v), \quad k=1,\dots, n,
    \label{HV1}
     \end{equation}
cf.~\cite{Huisman99,Uno}, where $x=(x_1, x_2,..., x_n)$ are the species abundances, $m\gg 1$, and $v=(v_1,..., v_n)$ the resource concentrations. Here $\mu_i$ are the species mortalities, $D_k >0$ are resource turnover rates, and $S_k$ is the supply of the resource $v_k$, $\phi_i$ is the specific growth rate of species as a function of the availability of the resource (also known as Michaelis–Menten's function). The coefficients $\gamma_{i} >0$ define self-limitation effects \cite{Roy}. We assume that each of the resources $v_k$, $k=1,\ldots,n$, is consumed by all species so that the content of $k$-th resource in the $i$-th species is positive $b_{ik} >0$.

We consider general $\phi_j$  which are bounded, non-negative and Lipshitz continuous
\begin{equation}
       \  0 \le \phi_j(v) \le  C_+, \quad  |\phi_j(v) -\phi_j(\tilde v)| \le L_j |v -\tilde v|,
\label{MM2a}
     \end{equation}
    i.e.,  $\phi_k$ have a minimal smoothness,  they are bounded and non-negative. The last restriction means that species consume
    resources.
     
 Moreover, we suppose 
\begin{equation}
      \phi_k(v) =0,   \quad  for \ all  \ k, \quad v \in \partial {\bf R}^m_{+}
\label{MM2b}
     \end{equation}
where $\partial {\bf R}^m_{+} $ denotes the boundary of the  hyperoctant $ {\bf R}_{+}^m =\{v:  v_j \ge 0, \ \forall j \}$.
Moreover, we suppose that
\begin{equation}
      \frac{\partial \phi_k(v)}{\partial v_j} \ge 0,   \quad  for \ all \ k,j,  \quad v \in \partial {\bf R}^M_{+}.
\label{MM5}
     \end{equation}
This assumption means that as the amount of the $j$-th resource increases all the functions $\phi_l$ also increase. 

Conditions~(\ref{MM2a}) and (\ref{MM2b}) can be interpreted as a generalization
of the well known von Liebig law, where
\begin{equation} \label{Liebm}
      \phi_k(v) =r_k\min \Big \{  \frac{ v_1}{K_{k1} +  v_1}, ...,  \frac{ v_m}{K_{km} +  v_m}\Big  \}
     \end{equation}
(cf.~\cite{Huisman99}) where $r_{k}$ and $K_{kj}$ are positive coefficients, and $k=1,..., M$. The coefficient
 $r_k$ is the maximal level of the resource consumption rate by the $k$-th species and coefficients $K_{ki}$, $i=1,..., M$ define the sharpness of the consumption curve $\phi_k(v)$.

A simple way to couple climate subsystem (\ref{sellers})
and the modeled biosphere defined by (\ref{HX1}) and (\ref{HV1})
is to suppose that the resource supply parameters
$S_k$ depends on the surface temperature $T$. Moreover,
we can suppose the albedo is a linear function of
$x_i$:
\begin{equation}
A=A(x)=A_0 + m^{-1} \sum_{j=1}^m c_j x_j.
\end{equation}
Finally, we obtain the following climate-biosphere system
\begin{equation}
     \frac{dx_i}{dt}=x_i (- \mu_i  + \phi_i(v) -  \gamma_{i} \; x_i), \quad i=1,\dots, m,
    \label{HX1M}
     \end{equation}
\begin{equation}
    \frac{dv_k}{dt}=D_k(S_k(T) -v_k)   -  \sum_{i=1}^m b_{ ki} \; x_i \; \phi_i(v), \quad k=1,\dots, n.
   \label{HV1M}
\end{equation}
\begin{equation} \label{sellers2M}
\frac{dT}{dt}=\lambda^{-1} \left(-e \sigma T^4 + \frac{\mu_0 I_0}{4}
\left(1- A_0 + m^{-1}\sum_{j=1}^m c_j x_j\right)\right).
\end{equation}

As an example, let us consider a model planet where the surface significantly covered by ice \cite{astrofeed} and the ice-albedo feedback is the main regulator of the planetary climate dynamics \cite{bioicefeed}.
Let the area of some region of the planet be $S_{arc}$, the area occupied by ice be $S_{ice}$ while the free ice area be $S_{free}$ \cite{SUD152}, where $S_{free}=S_{arc}-S_{ice}$. One can suppose that different species coexist in free ice domain and the averaged albedo of this domain is a linear combination of contributions of different species.
Then we obtain 
\begin{equation} \label{arct}
A_0=A_{ice} S_{ice}S_{arc}^{-1},  \quad c_j \propto S_{free}=S_{arc}-S_{ice},
\end{equation}
where $A_{ice}$ is the albedo of the ice-covered area. This relation will be useful below. 

Suppose that species populations $x_i$ and resources $v_k$ 
are fast variables, while the temperature
$T$ evolves slowly.  Such a situation arises
if, for example, $\gamma_i >>1$ (see 
\cite{Globstab}).  Then one can  show
that for large times $t$ 
$x_i(t) \approx X_i(T)$, where
$X_i(T)$ are time averaged equilibrium species populations for fixed
$T$ (see section \ref{sec:oursystem}). 
Then we obtain the following equation:
\begin{equation} \label{sellers2}
\frac{dT}{dt}=\lambda^{-1} \left(-e \sigma T^4 + \frac{\mu_0 I_0}{4}
\left(1- A_0 + m^{-1}\sum_{j=1}^m c_j X_j(T)\right)\right).
\end{equation}
Note that the equation (\ref{sellers2}) formally resembles the well-known ice-albedo feedback modification of the zero-dimensional energy balance model \cite{Sellers}. 

When the system (\ref{HX1M}), (\ref{HV1M}) and (\ref{sellers2})   
is regarded as a model of a particular biosphere, the choice of coefficients $c_k$ is determined by the environmental conditions at the given location and the corresponding species properties. Since we are aiming at building a global model, we want the eqs. 
(\ref{HX1M}), (\ref{HV1M}) and (\ref{sellers2}) to be applicable to any part of a modeled planet. Thus, we consider the coefficients unspecified. More precisely, we suppose that coefficients $c_k$ are random numbers described by certain probability distributions. We introduce these coefficients randomly assuming the randomness of the biological evolution.

In the coming section, we will consider a general class of slow-fast system with random coefficients, which includes the system (\ref{sellers2}) as a particular case.

\section{A general class of systems}\label{sec:general}
\subsection{A slow-fast  system}
In this section, we consider the following class of systems:
\begin{equation} \label{mainS1}
\frac{dy_i}{dt}=\kappa g_i(y, x),
\end{equation}
\begin{equation}\label{mainS2}
\frac{dx_j}{dt}=\sum_{l=1}^p
A_{jl} x_l + \kappa_1 F_j(y, x),
\end{equation}
where  $i=1, \ldots, n$,  $j=1, \ldots, p$,  and
$$
F_j(y, x)=\sum_{k=1}^m b_{jk} f_{k}(y, x).
$$
In these equations, the unknown vector-valued function $y(t)=(y_1(t), ..., y_n(t))$ consists of slow components, the unknown function $x=(x_1, ..., x_p)$ determines fast components, $\kappa, \kappa_1$ are
small positive parameters, $g_i, f_{k}$ are given smooth and uniformly bounded functions,
$b_{jk}$ are bounded coefficients,  and the square matrix $A_{jl}$  defines
a linear operator $A$ with  the spectrum $\sigma(A)$ such that
$$
Re \ \sigma(A) < -\delta_0 < 0.
$$
Then for sufficiently small $\kappa, \kappa_1>0$  the system of equations (\ref{mainS1}) and (\ref{mainS2}) has a locally attracting smooth and locally
invariant in an open neighborhood $U_{\kappa, \kappa_1}$ 
of $x=0$ manifold $\mathcal M$ defined by
\begin{equation}
\label{Inv}
x_j=\Phi_j(y, \kappa, \kappa_1)=  \kappa_1\Big(\sum_{k=1}^m c_{jk} f_{k}(y, 0))  +  \tilde X_j(y, \kappa, \kappa_1)\Big),
\end{equation}
where
$$
c_{ik}=-\sum_{j=1}^m (A^{-1})_{ij} b_{jk}.
$$
Here $A^{-1}$ stands for a matrix inverse to $A$ and
 sufficiently smooth functions $\tilde X_j(y, \kappa, \kappa_1)$  define small corrections such that
\begin{equation}
\label{InvtX}
 | \tilde X_j(\cdot, \kappa, \kappa_1)|_{C^1(U_{\kappa, \kappa_1})}
 \to 0  \quad (\kappa, \kappa_1 \to 0).
\end{equation}
Existence of $\mathcal M$  follows from the known results
(for example, \cite{He, Temam, Katok}).

As a result, we obtain the following system for slow variables:
\begin{equation}
\label{mainSF}
\frac{dy_i}{dt}=\kappa g_i(y, \Phi(y, \kappa, \kappa_1)),
\end{equation}
where $\Phi(y, \kappa, \kappa_1)=(\Phi_1(y, \kappa, \kappa_1), ..., \Phi_p(y, \kappa, \kappa_1))$.

For consideration of the systems with random parameters we need to use arguments from dynamical system theory and the Hoeffding inequality, one of concentration inequalities.

Recall the basic concept of structural stability introduced by A. Andronov and S. Pontryagin in 1937 \cite{Sm}.  Consider  a smooth vector field $F$ on  compact domain ${\mathbb D}^n$ of $\mathbb{R}^n$ with a smooth boundary  (or on a compact smooth manifold $M$ of dimension $n$). Assume that $F \in C^1({\mathbb D}^n)$ and consider all $\epsilon$-small perturbations
 $\tilde F$ such that
\begin{equation}\label{strucstab}
 |\tilde F|_{C^1({\mathbb D}^n)} <  \epsilon.
\end{equation}

Consider systems of differential equations $dx/dt=F(x)$ and $dx/dt=F(x) + \tilde F(x)$ and suppose that they define global semiflows $S_F^t$ and $S_{F +\tilde F}^t$ on ${\mathbb D}^n$. The system $dx/dt=F(x)$ is called structurally stable if there exists an $\epsilon_0$ such that for all positive $\epsilon < \epsilon_0$  trajectories of semiflows $S_F^t$ and $S_{F +\tilde F}^t$ are orbitally topologically conjugated (there exists a homeomorphism, which maps trajectories of the first system into trajectories of the second one). Roughly speaking, the original system is structurally stable if any sufficiently small $C^1$ perturbations of that system conserve the topological structure of its trajectories, for example, the equilibrium point stays an equilibrium (maybe, slightly shifted with respect to the equilibrium of non-perturbed system), or the perturbed cycle is again a cycle (maybe slightly deformed and shifted). We will refer the number $\epsilon_0(F)$
 the structural stability constant of the system 
$dx/dt=F(x)$.

Note that structurally stable dynamics may be, in a sense, "chaotic". There is a rather wide variation in different definitions of "chaos". Chaotic
(not periodic and no rest point) hyperbolic sets occur in some model systems
\cite{Katok, Sm,DH, NRT, Ruelle1, Ruelle2, Ruelle}.


\subsection{Systems with random parameters} \label{sec:5}

We consider systems (\ref{mainSF}), which arise, in a natural way, from systems decomposed in slow and fast variables.
We will use the following notation. We denote by $EX$ the expectation of a random quantity $X$, and by $Var \ X$ its variance. Moreover, $\Pr[A]$ denotes the probability of a random event $A$. In this section, we formulate general principles on averaging with respect to the parameters that are applicable to fast-slow climate models.


Consider the following general system of differential equations:

\begin{equation}\label{GCS}
\frac{dy_i}{dt}= g_i(y, \Phi (y)),
\end{equation}
where  $i=1, \ldots n$,
$y(t)=(y_1(t),\ldots  y_n(t))$ is a unknown vector-function, and $\Phi=(\Phi_1, ... , \Phi_p)$,
$\Phi_l(y)$ are functions, which will be defined below.  Let ${\mathbb B}^n$ be a compact subdomain of $\mathbb{R}^n$  with a smooth boundary $\partial {\mathbb B}^n$. We suppose
that  $g_i(y, \Phi)$ are smooth functions uniformly bounded as are the first and second derivatives with respect to all variables $y, \Phi$:
 \begin{equation}\label{aprioriF}
|g_i|_{C^2({\mathbb B}^n \times {\mathbb R}^p)} < C_g,
\end{equation}
where $C_g$ is a positive constant.

We assume, moreover, that the functions $\Phi_l(y)$ are linear combinations of other functions $f_{j}(y)$ with random coefficients $c_{ij}$:
\begin{equation}\label{DF}
\Phi_i(y)=m^{-1}\sum_{j=1}^{m}  c_{ij} f_{j}(y),
\end{equation}

We suppose that the $f_{j}$ are non-random, fixed functions and they have uniformly bounded  derivatives
\begin{equation}\label{Derf}
  |f_{j}| _{C^2({\mathbb B}^n)} < C_f,
\end{equation}
where a positive constant $C_f$ is uniform in $i, j$.

For (\ref{GCS}) we set the initial data
\begin{equation}\label{initdata}
y(0)=y^{(0)}.
\end{equation}


Let the following assumptions hold:

\begin{ass} 
{\em Let $c_{ij}$ be independent  random quantities such that
$Ec_{ij}=\bar c$, and, moreover, almost surely
\begin{displaymath}
c_{ij} \in (-R_0, R_0),
\end{displaymath}\label{cpdf}
where $R_0 >0$ is a constant uniform in $m$.}
\end{ass}

As a consequence, one has  that if
\begin{displaymath}
E (c_{ij} - \bar c)(c_{kl} -\bar c) = 0 \quad if \quad  i \neq k \ or \ j \neq l.
\end{displaymath}\label{Cind}

Together with system (\ref{GCS}) we consider the corresponding averaged system:
\begin{equation}\label{GCSA}
\frac{d\bar y_i}{dt}=\bar g_i(y),
\end{equation}
where
\begin{equation}\label{barF}
\bar g_i(y)= g_i(\bar y, \bar \Phi_1(  y), ..., \bar \Phi_p(  y) ),
\end{equation}
where $i=1, \ldots,  n$, and
$ y(t)=( y_1(t),\ldots   y_n(t))$ is a unknown vector-function,   and
$\bar \Phi_i(  y)$ are averages of  functions $\Phi_i(  y)$ over the random parameters $c_{ij}$:
\begin{equation}\label{DFE}
\bar \Phi_i( y)=  \bar c  m^{-1} \sum_{j=1}^m  f_{j}(y) .
\end{equation}

We assume that there hold the following conditions:
\begin{equation}\label{towards}
\bar g(y) \cdot e(y) < 0 \quad  \forall y \in \partial {\mathbb B}^n,
\end{equation}
and
\begin{equation}\label{towards1}
 g(y, \Phi(y)) \cdot e(y) < 0 \quad  \forall y \in \partial {\mathbb B}^n,
\end{equation}
where $e(y)$ is a normal vector to the boundary $\partial {\mathbb B}^n$ at the point $y$ directed inward on the domain ${\mathbb B}^n$. For the system (\ref{GCSA}) we set the same initial data (\ref{initdata}). Condition (\ref{towards}) implies that the Cauchy problem (\ref{initdata}) and (\ref{GCSA}) defines a global semiflow on the domain ${\mathbb B}^n$.

\subsection{Main features of the systems with random parameters}
For slow variables systems (\ref{mainSF}) we prove an averaging theorem assuming that $c_{ik}$ are random independent parameters (see the Appendix).  This theorem asserts the attractor of the original system is close to the attractor of averaged one with a probability $Pr_m$, which is exponentially close to $1$ for large $m$. So, our main idea in the explanation the relative stability of climate is that a large number of independent factor can mutually cancel each other out. 
The probability $Pr_m$ satisfies an inequality that includes the number $\epsilon_0$ is a measure of stability under perturbations. If $\epsilon_0>0$ is small,
 i.e., the original system is weakly stable and conserves its dynamics only under very small perturbations, then estimate
 (\ref{EstDiffpr}) makes a sense only for large $m> m_0(\epsilon_0)$ (in fact, for bounded $m$ the right hand side of (\ref{EstDiffpr}) is negative). 
 
 Moreover, the structurally stable system are seldom found in real applications
(if we exclude the cases $n=1$ and $n=2$, where they are generic).  According to basic result of S. Smale \cite{Katok, Ruelle}, for  dimensions $n >2$ structurally stable systems are not generic. To overcome this difficulty, we consider an approach, which allows us to show that solutions of the original system stay in a small neighborhood of a local attractor of the corresponding averaged system.

The stability of many dynamical regimes can be proved by using Lyapunov functions.
Recall that $L(y)$ is a Lyapunov function of a system $dy/dt=g(y)$  in a domain ${\mathbb V} \subset {\mathbb R}^n$ if $L$ is at least $C^1$ smooth and $L(y(t))$ does not increases along trajectories $y(t)$ of the system:
\begin{equation} \label{LF} \nabla L(y) \cdot g(y) \le 0, \quad y \in {\mathbb V}.
\end{equation}

For example, if $y^*$ is a stable rest point of the system,
then one can construct a $L(y)$ close to a quadratic
form, which is Lyapunov function in a small neighborhood ${\mathbb V}$ of $y^*$ and
\begin{equation} \label{LF1}
 \nabla L(y) \cdot g(y) \le c|y-y^*|^2, \quad y \in {\mathbb V}\end{equation}
for some $c >0$.

The next statement (see the Appendix) can be proven for the Lyapunov functions.  If the averaged system defined by (\ref{GCSA}) has a Lyapunov function then  the original system (\ref{GCS}) has {\em the same} Lyapunov function a  probability $Pr_L$, which is exponentially close to $1$  as $m$ large.

This theorem can be applied to the Budyko--Sellers energy balance system (\ref{sellers2}) as follows. Suppose that the averaged system is gradient-like (note that (\ref{sellers2}) enjoys this property). Let $\bar A$ be an attractor of the original system, which  consists of stable equilibria. Suppose that all equilibria of the averaged system are hyperbolic. Then there exists a Lyapunov function $L_(y)$ such that
$$
H_{\bar g} (y)= \nabla L(y) \cdot \bar g(y) \le -\epsilon,
$$
for all $y \in {\mathbb V}(\bar A)  $ and some $\epsilon >0 $, where ${\mathbb V}(\bar A)$  is an open subset of the attraction basins
of $\bar A$. This subset contains all points $y$ except for
small $\delta$-neighborhoods of equilibria, where
$\delta \to $ as $\epsilon \to 0$.
Then with probability $Pr_{\delta, \epsilon, m}$  all original
system also has the same Lyapunov function with
analogous properties. 

\section{Stability of the coupled climate-biosphere system}\label{sec:oursystem}

Let us apply (\ref{mainT2}) and
(\ref{mainTL}) to a system defined by (\ref{sellers2}). In the general case this system is complicated. To simplify the problem, we suppose that the $c_i$ are random independent quantities such
that $E c_i=\bar c$, and, moreover, we apply the approximation obtained in \cite{Uno, sud16,vak18}. We assume that the turnovers satisfy
$D_k >>1$.  Then
$$
v_k=S_k - \tilde S_k, \quad 0 < \tilde S_k < const D^{-1}.
$$
Suppose that all species $X_j$ survive and have positive
abundances. Then
$$
X_j(T)=U_j(T) + O(D^{-1}),
$$
$$
U_j(T):=\gamma_i^{-1}(\phi_i(S(T)) -\mu_i)_{+},
$$
where we use notation $f_{+}=max(f, 0)$.
Then eq. (\ref{sellers2}) take the form (we remove the terms the order $O(D^{-1})$)
\begin{equation} \label{sellers3}
\frac{dT}{dt}=\lambda^{-1} \big(-e \sigma T^4 + \frac{\mu_0 I_0}{4}
(1- A_0 + m^{-1}\sum_{j=1}^m c_j U_j(T)) \big).
\end{equation}
We apply (\ref{mainT2}) and (\ref{mainTL}), with $p=1$ and
$$
\Phi_1=m^{-1}\sum_{j=1}^m c_j U_j.
$$
The averaged system takes the form
\begin{equation} \label{sellers34}
\frac{dT}{dt}=\lambda^{-1} \Big(-e \sigma T^4 + \frac{\mu_0 I_0}{4}
\big(1- A_0 + C w(T)\big)\Big),
\end{equation}
where
$$
w(T)=m^{-1} \sum_{j=1}^m  U_j(T), \quad C=m^{-1}\sum_{i=1}^m E c_i=\bar c.
$$
Let all $\phi_i(S)$ be uniformly bounded by a constant $a$, $\phi_i(S) < a$ for
all $i=1,...,m$ and $S$.
Then we find that, with a probability exponentially close
to $1$, there exists a Lyapunov function defined by
$$
L(T)= - \frac{e \sigma T^5}{5} + \frac{\mu_0 I_0}{4}\big( (1- A_0) T + C W(T)\big),
$$
where
$$
W(T)=\int_0^T w(s) ds.
$$
Non-degenerate local minima of this function are steady states (local attractors) of the averaged system, and local extrema are saddle points or repellers of that system. If $\bar c$  is small enough, we have only a single local attractor $T=\bar T_e$.
Our theorems (see the Appendix) imply that the original system then also has
(with a probability close to $1$) a single local attractor $T=T_e(m)$ and $|T_e(m)-\bar T_e| \to 0$ as $m \to \infty$.

The situation  dramatically changes if the condition $\phi_i < a$ is violated, say, one species dominates or if $m$ is small.  Then it is impossible to guarantee that $|T_e(m)-\bar T_e| \to 0$. This means that  biodiversity decreases  can produce global climate changes.

To find possible bifurcations, we consider the simplest case when we are dealing with a single resource $v_1=v$ and the growth function are identical for all species, 
$\phi_i(v)=v(K_i+v)^{-1}$. We assume that 
$S(T)=S_0 + S_1\Delta(T)$, where the coefficient $S_1$ defines
an influence of temperature on the resource supply and
$$
\Delta(T)=\exp(- (T- T_0)^2/2\sigma_T^2)
$$
This means that there exist an optimal 
temperature $T_0$ for species growth and
a characteristic spread of this temperature $\sigma_T$.
Then we obtain
eq. (\ref{sellers34}) with
$$
w(T)= m^{-1} \sum_{i=1}^m \frac{S_0 + S_1\Delta_T}{K_i + S_0 + S_1\Delta_T}
$$
and the equation for temperature steady state takes then the form
\begin{equation} \label{maineq}
F(T)=G(T),
\end{equation}
where 
$$
F(T)=e \sigma T^4, \quad G(T)=\mu_0 \frac{I_0}{4} (1- A_0 + C w(T)).
$$
Note that for $C >0 $ the species diminish the averaged planetary albedo. Moreover, simulations show that for small $m$ variations in the species parameters, for example, in $K_i$ can decrease albedo.

\begin{figure}
 \includegraphics[scale=0.8]{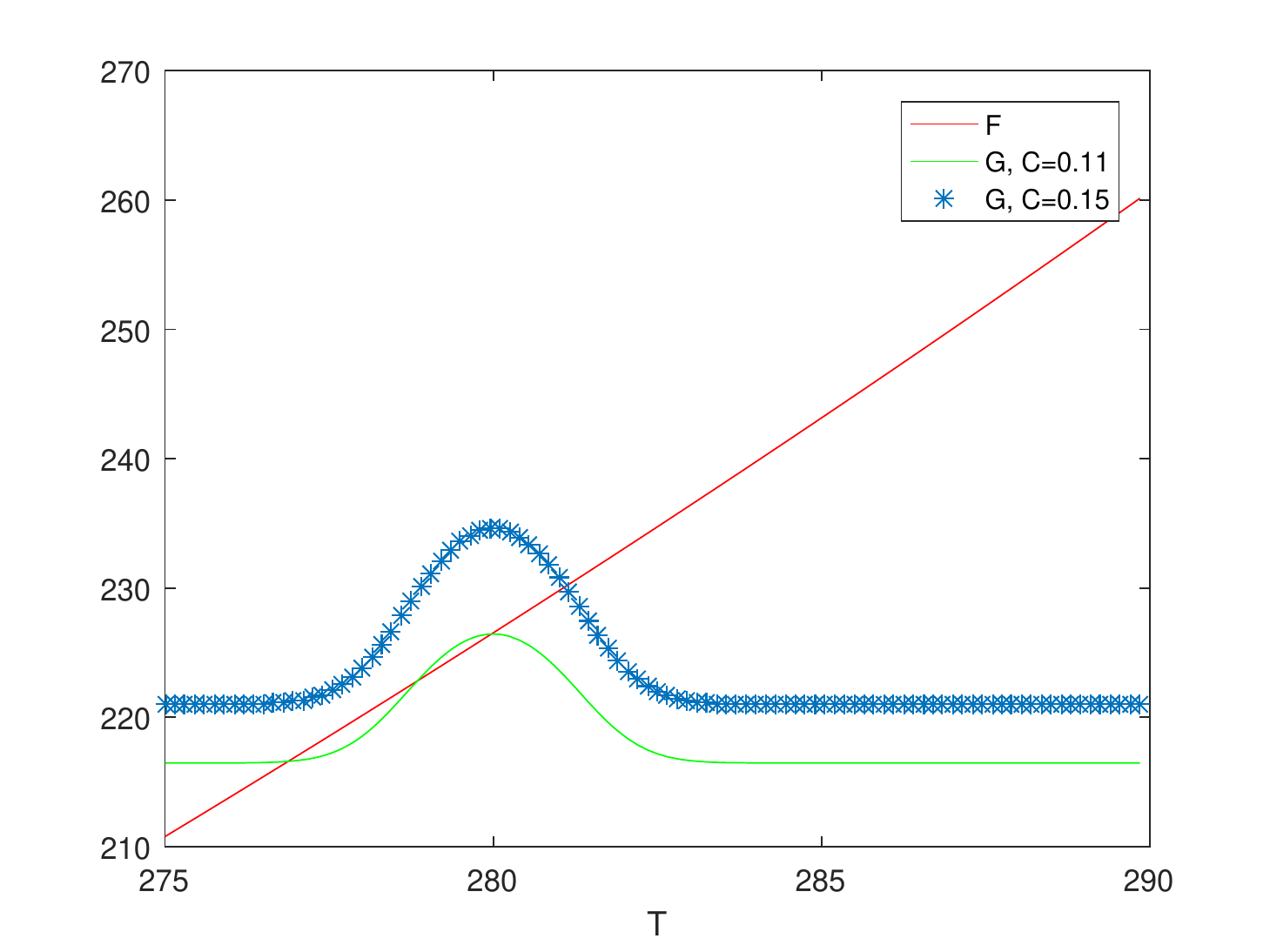}
\caption{\small This plot shows possible bifurcations in climate-biosphere system. The equilibrium temperature values are given by intersections of curves $F(T)$ and $G(T)$. We have a single intersection for  $C=0.15$ and the three  intersections for  $C=0.11$. For the biosphere, we have $m=5$ species, where the parameter values are $K_i=0.1, S_0=0.1, \mu=0, S_1=0.2, T_0=280K$ and $\sigma_T=1$. We use the parameters similar to the Earth's climate system, we have $\sigma=5.67\cdot 10^{-8}, A_0=0.62, \mu_0=1, e=0.65,I_0/4=340$ and $\bar c=0.2$.}
\label{Bif}
\end{figure}

Depending on $C$ we have either a single root of (\ref{maineq}), or three, when two roots give us local attractors and the third  root is a saddle point,  as it is shown on Fig. \ref{Bif}.  So, we observe here a pitchfork bifurcation, which is  essentially the same as for the ice-albedo feedback problem, see
above.

It is interesting to understand as global warming affects the described bifurcation effect. Consider the cold planetary region and relation (\ref{arct}).  We observe that a decrease of the area occupied by ice increase the coefficient $\bar c$ and decreases $A_0$ thus it reinforces the bifurcation effect and can lead to climate bifurcation. 

\section{Discussion and conclusions} \label{sec: 13}
Understanding the planetary climate dynamics and identification of factors and processes that can affect its stability are problems of significant importance, in particular, because of their prominent effect on the biosphere functioning. There is growing evidence that the biosphere can have a variety of feedback loops to climate and a comprehensive understanding is only possible based on the analysis of the coupled climate-biosphere system. For example, in the Earth system, the perturbation of the carbon cycle \cite{sudakov13} or water-vapor \cite{vapor} cycle is one such feedback, but there are many more. Here we focus on the feedback induced by an interaction between the biosphere of an ice planet and climate.
In this paper, we have endeavored to address this issue theoretically, in the framework of the Gaia hypothesis by considering a conceptual model of climate-biosphere dynamics arising from the coupling between a global energy balance model to a generic multi-specific model of population dynamics. 

The climate-biosphere system is an extremely complex system and the corresponding mathematical model, even a relatively simple 'conceptual' one, is usually too complicated for a comprehensive analytical study. A possibility of nontrivial model reduction lays in the observation that different processes often go with very different rates, i.e.~take place on very different timescales. In particular, many complex systems, including climate models, have slow and fast components.  According to classical results \cite{CFNT},  large time dynamics of fast modes are captured by a dynamics of  slow modes on a slow invariant manifold. It is well known that even low dimensional systems exhibit complex bifurcations \cite{Engler, Salt1,sieber18}. Moreover, such models exhibit multistationarity, i.e., existence of many stationary states that, according to \cite{Emanuel}, provides the climate stability under variations of astronomical factors.

\begin{figure}
\includegraphics[scale=0.45]{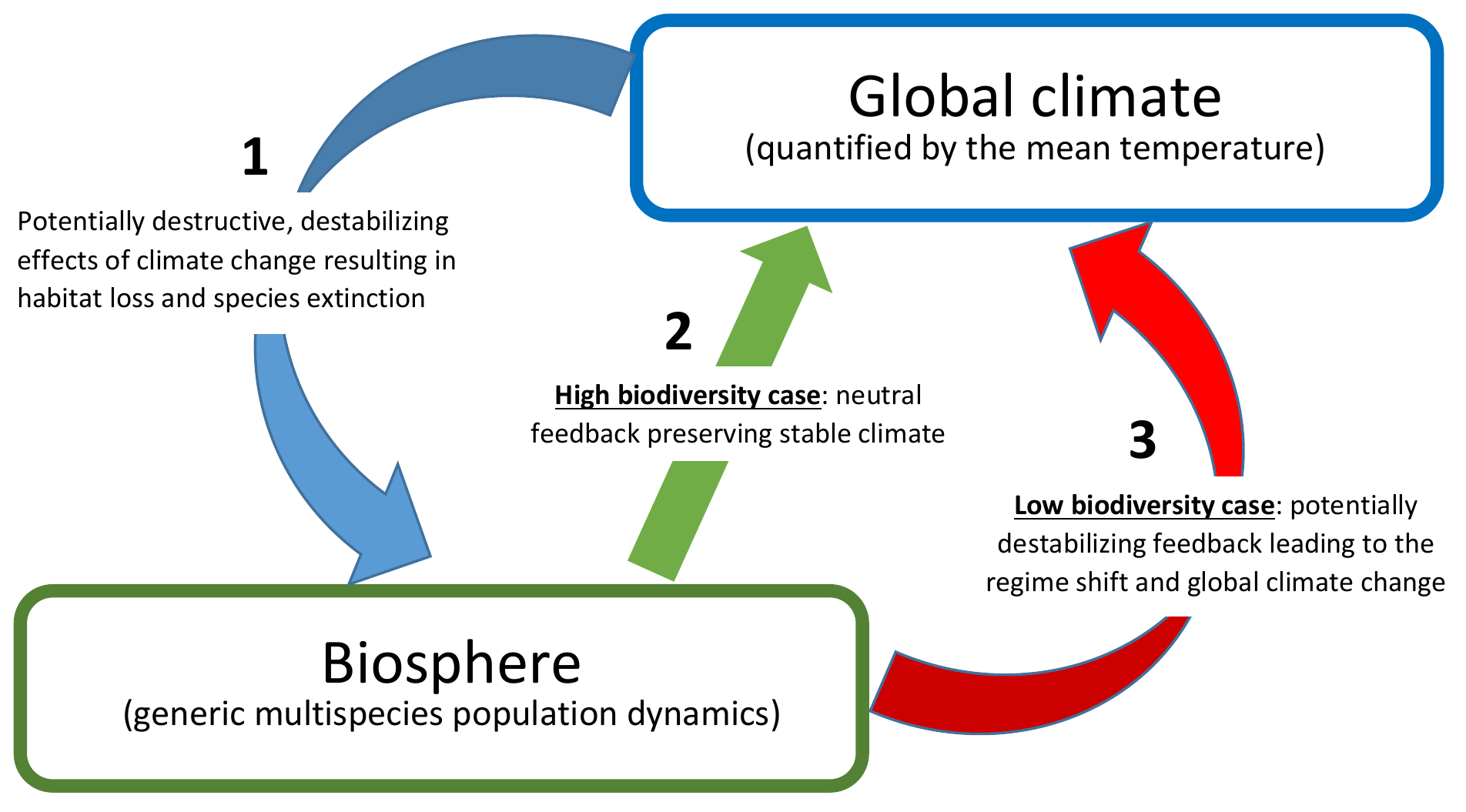}
\caption{\small Graphical summary of the feedbacks in our model coupled climate-biosphere system (\ref{HX1M}, \ref{HV1M} and (\ref{sellers2M})). Arrow 1 shows the potentially destructive effect of the global climate change on the population dynamics and ecosystems functioning. Arrow 2 shows the neutral feedback that the population dynamics have on the global climate in case of high biodiversity, i.e.~a large number of coexisting species). Arrow 3 shows the potentially destabilizing feedback of the population dynamics on the global climate in case of low biodiversity.}
 \label{diagram}
\end{figure}


Referring back the Gaia concept, why, however, was the climate system stable over long periods of time in the past? To answer this question, we assumed that parameters of fast subsystems are random and mutually independent.  Under such assumptions, we prove a general theorem on connection between attractors of averaged and original systems. If the attractor $\bar A$ of the averaged system has a low fractal dimension then, with a probability close to $1$, the attractor of the original system is close to $\bar A$. We think that this result may have applications for  many different fields such as global network systems with unknown parameters, foodwebs, gene networks etc.

So, climate stability can be explained by the fact that many
independent factors are canceled out.  In our realization of the Gaia concept, the stability of the climate system is ensured by growing biodiversity.
Our findings are summarized in Fig.~\ref{diagram}. Interestingly, our analysis suggests a possibility of a positive feedback of the biosphere on the climate change. Consider a scenario of a slow change in the energy balance resulting, for instance, in a gradual increase of the mean temperature. Its is well known that such an increase will eventually results in species extinctions and biodiversity loss (see Arrow 1 in Fig.~\ref{diagram}). Our results predict that, as long as the number of extinctions is not too large, the biodiversity loss will not have any notable feedback on the climate dynamics (Arrow 2 in Fig.~\ref{diagram}). However, when the biodiversity loss becomes considerable, i.e.~the number of surviving species becomes small, the failing biota will have a positive feedback on the climate resulting in its destabilization (Arrow 3 in Fig.~\ref{diagram}). The global climate change resulting from this destabilization is likely to have a stronger negative effect on the biosphere, hence accelerating the extinction rate.

This model may be used to reconstruct and project climate change on the ice planets of the Solar System \cite{Icarus,astrobio} and some exoplanets \cite{Natgeo}. Another possible application of our approach is paleoclimate modeling. For example, in the Cryogenian period, the planet has been transformed into so-called ‘snowball Earth’, where early life survived under the environmental stress \cite{Neoprot}, became stable and even diverse \cite{alage}. Our model may help to evaluate how biodiversity could contribute to global ice melting in this period. 


 
\section*{Acknowledgments} 
This research is supported by the Swedish Foundation for International Cooperation in Research and Higher Education (STINT), Grant IB 2018-7517. 
This work was assisted by attendance as a Short-term Visitor at the National Institute for Mathematical and Biological Synthesis, an Institute supported by the National Science Foundation through NSF Award DBI-1300426, with additional support from The University of Tennessee, Knoxville. We also would like to thank the Mathematical Biosciences Institute (MBI) at Ohio State University, for helping initiate this research. MBI receives its funding through the NSF grant DMS-1440386. IS and SVP acknowledge the kind hospitality of Banff International Research Station (BIRS) for Mathematical Innovation and Discovery where they worked on this research. DL has been supported by the Government of the Russian Federation (the grant 08-08). IS also thanks University of Dayton Research Council Seed Grant, 2019.
\section*{Appendix}
\label{Appendix}

In this Appendix,
constants $c$ and $C_i$ can depend on system parameters but are uniform in $m$ for large $m$. Note that we sometimes denote
different constants by the same index if it does not lead to confusion. Our proving plan can be outlined as follows. To simplify our statement, we first prove three auxiliary lemmas, and then we state short
demonstrations of theorems.

{\em Probabilistic estimates}.
Let us fix some points $y^{(k)} \in {\mathbb B}^n$, where $k=1, 2,..., M$ and $M$ is an positive integer, which will be adjusted later.
Let us define the events ${\mathcal A}_{\epsilon, i}(k)$ by
\begin{equation}\label{events}
{\mathcal A}_{out,\epsilon, i }(k)= \{ |\bar g_i (y^{(k)}) -   g_i(y^{(k)}, \Phi(y^{(k)}))| > \epsilon/4  \},
\end{equation}
\begin{equation}\label{events1}
{\mathcal A}_{\epsilon, i}(k) =Not  \   {\mathcal A}_{out, \epsilon, i}(k),
\end{equation}
where $Not \ B$ denotes the negation of $B$ {\em and $\bar g_i(y)$ are defined by relation (\ref{barF})}.

The next auxiliary lemma is elementary but useful.

\begin{lem} \label{canal}
{ One has
$$
\Pr \big [ \prod_{k=1}^M \prod_{i=1}^n  {\mathcal A}_{\epsilon, i}(k) \big] \ge  1 - \sum_{k=1}^M
\sum_{i=1}^n \Pr \big [ {\mathcal A}_{out,\epsilon, i }(k) \big].
$$
}
\end{lem}
{\bf Proof.} That lemma can be proved by de Morgan's rule.


Furthermore, we use Chernoff  bounds to estimate $\Pr \big [ {\mathcal A}_{out,\epsilon, i }(k)) \big]$. Let $C_{\bar g, \Phi}$ be a Lipshitz  constant of $\bar g$ with respect
to the variables $\Phi_1, ..., \Phi_p$, i.e., for all
$y \in {\mathbb B}^n$ and $i=1, ..., n$
\begin{equation} \label{Lipg}
|\bar g_i(y, \Phi^{(1)}) - \bar g_i(y, \Phi^{(2)})|
\le C_{\bar g, \Phi} |\Phi^{(1)} - \Phi^{(2)}|, 
\end{equation}
where   $|\Phi|=\max_l \ |\Phi_l|$. This constant $C_{\bar g, \Phi}$ exists
due to assumption (\ref{aprioriF}) to $g$.
Moreover, an analogous estimate holds for derivatives with respect to $y$:
all $i,j$: 
\begin{equation} \label{LipgD}
|\nabla_y \bar g_i(y, \Phi^{(1))}) - \nabla_y \bar g_i(y, \Phi^{(2)})|
\le \tilde C_{\bar g, \Phi} |\Phi^{(1)} - \Phi^{(2)}|. 
\end{equation}

\begin{lem} \label{Chern}
{ One has
\begin{widetext}
\begin{displaymath}\label{Chern1}
\Pr \big [ {\mathcal A}_{out,\epsilon, i }(k) \big] < 2p \exp\big(-  m \epsilon^2/(8 C_{\bar g, \Phi}^2 C^2)\big), \quad \forall \ i=1,.., n, \ k=1,..., M,
\end{displaymath}
\end{widetext}
where
$$
C= R_0 \max_{ j, k}  (|f_{j}(y^{(k)})| + |\nabla_y  f_{j}(y^{(k)}|).
$$
}
\end{lem}
{\bf Proof.} Our the first step is to estimate differences $\Phi_i(y^{(k)}) - E\Phi_i(y^{(k)})$. To this end, let us fix  indices $i$ and $k$ and introduce $X_j$ by
\begin{equation} \label{Xjy}
X_j =c_{ij} f_{j}( y^{(k)}).     
\end{equation}
Then
\begin{equation} \label{Pjy}
\Phi_i(y^{(k)}) =m^{-1} \sum_{j=1}^m X_j.     
\end{equation}
 Our {\bf Assumption \ref{cpdf}} on $c_{ij}$ implies
that $X_j$ are independent random variables. Moreover, taking into account
that $C^1$ - norms of $f_{j}$ are uniformly bounded we have
\begin{equation} \label{Xui}
|X_j|   < C.
\end{equation}

Let us recall the Hoeffding inequality. Let $X_j$, $j=1,..., m$ be  independent random variables strictly bounded in intervals $[a_j, b_j]$, i.e. surely $X_i \in [a_i, b_i]$. Let $\bar X=m^{-1} \sum_{j=1}^m X_j$ be the average of those quantities. Then (see \cite{Hoeff})
$$
\Pr[ |\bar X - E\bar X| \ge t) \le 2  \exp\Big( -\frac{2m^2 t}
{\sum_{j=1}^m (a_i -b_i)^2 }\Big).
$$
Therefore, according to Hoeffding's inequality 
 for each $\epsilon>0$ we obtain
\begin{equation} \label{exp66f}
{\Pr}[ |\Phi_l(y^{(k)})-  E\Phi_l(y^{(k)})| > \epsilon]  < 2 \exp( -2 m \epsilon^{2}/C^2),
\end{equation}
where $l=1,..., p$. 

The second step is as follows. 
Consider the events
$$
{\mathcal B}_{\delta, l, k}=\{|\Phi_l(y^{(k)}) -  E\Phi_l(y^{(k)})|  < \delta \}.
$$
Let ${\mathcal B}_{\delta}=\prod_{l=1}^p  {\mathcal B}_{\epsilon, l, k}$
Then, due to (\ref{canal}) and (\ref{exp66f}),
 \begin{equation}\label{lastest}
\Pr \big [ {\mathcal B}_{\delta} \big] \ge   1- 2p \exp\big( - 2 m \epsilon^2/C^2 \big).
\end{equation}
Let us take $\delta=\epsilon/4 C_{\bar g, \Phi}^2$. Then, if the event ${\mathcal B}_{\delta}$ takes place, we have  
(because  $g$ is a Lipshitz map with the Lipshitz constant $C_{\bar g, \Phi})$ and by definition of $\bar g$) that
$$
|\bar g(y^{(k)}) - g(y^{(k)}, \Phi(y^{(k)}))| < \epsilon/4,
$$
i.e., the event opposite to $A_{out,\epsilon, i }(k)$ takes place.
Now   conditions (\ref{aprioriF}) and  estimate (\ref{lastest}) lead to  inequality (\ref{Chern1})
that completes the proof of the lemma.

Let us define now the events ${\mathcal A}_{out,\epsilon, i, j }(k)$ and ${\mathcal A}_{\epsilon, i, j}(k)$ by
\begin{equation}\label{eventsD}
{\mathcal A}_{out,\epsilon, i, j }(k)= \{ | g_{ij}(y^{(k)}) -   g_{ij}(y^{(k)})| > \epsilon/4n  \},
\end{equation}
where
$$
\bar g_{ij}(y)=\frac{\partial \bar g_i (y)}{\partial y_j},
\quad
g_{ij}(y)=\frac{\partial g_i(y, \Phi(y))}{\partial y_j},
$$
and
\begin{equation}\label{events1D}
{\mathcal A}_{\epsilon, i,j}(k) =Not  \   {\mathcal A}_{out, \epsilon, i}(k).
\end{equation}

There holds the following Lemma:
\begin{lem} \label{Chern2}
{One has
\begin{widetext}
\begin{equation}\label{Chern1a}
\Pr \big [ {\mathcal A}_{out,\epsilon, i,j }(k) \big] \le 2p \exp(-  m\epsilon^2/(C^2 \tilde C_{\bar g,\Phi}^2) ), \quad \forall i, j=1,.., n, \ k=1,..., M,
\end{equation}
\end{widetext}
where $\tilde C$ is defined by (\ref{LipgD}).
}
\end{lem}

The proof of (\ref{Chern2}) repeats the same arguments used in the proof of (\ref{Chern}) so do not present it.

\subsection*{Demonstrations of Theorem \ref{mainT2} and Theorem \ref{mainTL}}
\begin{thm}  \label{mainT2}
{Suppose condition (\ref{towards}) holds and that averaged system defined by (\ref{GCSA}) defines a global dissipative semiflow on the  domain ${\mathbb B}^n$. Moreover, let us assume that averaged system
(\ref{GCSA}) is structurally  stable with 
a  structural stability constant $\epsilon_0(\bar g)$ and that system has an  attractor $\bar A$. Then  with probability
 $\Pr_{\bar A}$
the original system (\ref{GCS}) also defines a global dissipative semiflow on ${ \mathbb B}^n$, which has an attractor
$A$  topologically equivalent to $\bar A$. The probability
 $\Pr_{\bar A}$ satisfies the inequality
\begin{displaymath}\label{EstDiffpr}
Pr_{\bar A} > 1-  C_1 n  \exp\big (- C_2 m \epsilon_0^{2}  - n \ln \epsilon_0 \big),
\end{displaymath}
where $ C_1, C_2$ are  positive constants uniform in $m$}.
\end{thm}
{\bf Proof.} We use (\ref{canal}), (\ref{Chern}) and (\ref{Chern2}) and the following auxiliary construction. The domain
${\mathbb B}^n$ has the dimension $n$ therefore we can cover it by $N(r \epsilon) \sim
 (r\epsilon)^{-n}$ balls $\Omega_{\epsilon,k}$ of the radius $\epsilon$ centered at some points $y^{(k)} \in {\mathbb B}^n $. Here $r$ is a positive constant uniform in $\epsilon$.
We denote the union of all  those balls  by $U_ {\epsilon}$, it is an open neighborhood of ${\mathbb B}^n$.

Let us consider the perturbation $\tilde g(y)=g(y, \Phi(y)) - \bar g(y)$ and  estimate the $C^1$ norm of $\tilde g$ on $U_ {\epsilon}$.  Suppose that all events ${\mathcal A}_{\epsilon, i}(k)$ and ${\mathcal A}_{\epsilon, i,j}(k)$
defined by (\ref{events1}) and
(\ref{events1D}), respectively,  take place.
Then
\begin{equation} \label{coverest}
|\tilde g(y^{(k)})|  +  |\nabla_y \tilde g(y^{(k)})|  < \epsilon/2, \quad k=1,...,  N(\epsilon).
\end{equation}
Then, due to conditions (\ref{aprioriF}) on $g$, and definition of $\bar g$ we have
$$
|\tilde g|_{C^2({\mathbb B^n})} < C_1,
$$
    where a positive constant $C_1$ is independent of $m$. 
    Therefore, for each $y \in {\mathbb B}^n$
    one can find such point $y^{(k)}$ that 
    there hold the estimates
    $$
    |\tilde g_i(y^{(k)}) - \tilde g_i(y)| < r\epsilon,
    $$
    $$
    | \frac{\partial \tilde g_i(y^{(k)})}{
    \partial y_j} - \frac{\partial \tilde g_i(y)}
    {\partial y_j}| < r\epsilon.
    $$
 Those last inequalities and (\ref{coverest}) imply
\begin{equation} \label{coverest2}
|\tilde g(y)|  +  |\nabla_y \tilde g(y)|  < \epsilon/2 + C_2 r\epsilon,  \quad y \in U_ {\epsilon},
\end{equation}
where $C_2$  is a positive constant. We set $r= 1/2C_2$. 
Due to conditions (\ref{towards}) and  (\ref{towards1}) the vector fields $g$ and $\bar g$ are
directed towards interior of ${\mathbb B}^n$
that allows us to apply now the definition of structural stability \cite{Ruelle}.
 Then for positive $\epsilon \le \epsilon_0(\bar g)$ the attractor of
 the original system is topologically equivalent to the attractor of the averaged system. Note that  $\epsilon_0$
 does not depend on $m$ and it is defined by the averaged system only.

Furthermore, we compute the probability that all the events defined by (\ref{coverest}) take place by (\ref{canal}), (\ref{Chern}) and (\ref{Chern2}).
This finishes the proof.

\begin{thm}  \label{mainTL}
Suppose condition (\ref{towards}) holds and that the averaged system defined by (\ref{GCSA}) has a Lyapunov function
such that
\begin{equation} \label{LFeps}
 \nabla L(y) \cdot \bar g(y) \le -\epsilon, \quad y \in {\mathbb V}
\end{equation}
where ${\mathbb V}$ is an open subdomain of ${\mathbb R}^n$
with a compact closure, and moreover,
$$
|L|_{C^2( {\mathbb V})} < C_L
$$
for a positive constant $C_L$.
Then    with the probability
 $\Pr_{L, \epsilon}$
the original system (\ref{GCS}) has {the same} Lyapunov function such that
\begin{equation} \label{LFeps2}
 \nabla L(y) \cdot g(y) \le -\epsilon/2, \quad y \in {\mathbb V}.
\end{equation}
The probability
 $\Pr_{L, \epsilon}$ satisfies the inequality
\begin{displaymath}
\label{EstDiffpr1}
Pr_{L, \epsilon} > 1-  \bar C_1   \exp\big (- \bar C_2 m \epsilon^{2}  -  \ln \epsilon \big),
\end{displaymath}
where $ \bar C_1,  \bar C_2$ are  positive constants uniform in $m$.
\end{thm}

Let us note that, similarly to the previous theorem,   If $\epsilon>0$ is small,
  estimate
 (\ref{EstDiffpr1}) makes a sense only for large $m> m_0(\epsilon)$.

{\bf Proof.} We apply the same idea used in the previous proof.
 The domain
${\mathbb V}$ can be  covered by $N(r \epsilon) \sim
 (r\epsilon)^{-n}$ balls $\Omega_{\epsilon,k}$ of the radius $\epsilon$ centered at some points $y^{(k)} \in 
 {\mathbb B}^n$. Here $r$ is a positive constant uniform in $\epsilon$. Let us introduce the functions
$$
   \bar H(y)=\nabla_y L(y) \cdot \bar g(y),
\quad
    H(y)=\nabla_y L(y) \cdot  g(y, \Phi(y)).
$$
Consider the events
\begin{equation}\label{eventH}
{\mathcal H}_{out,\epsilon}(k)= \{ |H (y^{(k)}) -   \bar H(y^{(k)})| > \epsilon/4  \},
\end{equation}
\begin{equation}\label{outH}
{\mathcal H}_{\epsilon}(k) =Not  \   {\mathcal H}_{out, \epsilon}(k)=\{ |H (y^{(k)}) -   \bar H(y^{(k)})| \le \epsilon/4  \}.
\end{equation}
Suppose that all events defined by (\ref{outH}) take place.
Then
\begin{equation} \label{coverestH}
|H(y^{(k)}) - \bar H(y^{(k)})|\ < \epsilon/4, \quad \forall \ k=1,...,  N(\epsilon).
\end{equation}
Now we use the estimate
\begin{equation} \label{gradtH}
|H(y^{(k)})- H(y) | < Lip_{H} |y^{(k)} -y |,
\end{equation}
where $Lip_{H}$ is a Lipshitz constant of $H$. Let us estimate
that constant. By definition of $H$ one has
$$
\frac{\partial  H}{\partial y_k}= m^{-1} \sum_{i=1}^n \sum_{j=1}^m
c_{ij} \frac{\partial ( L f_{j})}{\partial y_k}.
$$
Due to {\bf Assumption \ref{cpdf}}
\begin{equation}
\big | \sum_{i=1}^n \sum_{j=1}^m
c_{ij} \frac{\partial ( L f_{j})}{\partial y_k}  \big| < m n c_1
R_0,  
\end{equation}
where
\begin{equation}
c_1=\max_{i, j, y \in {\mathbb V} } ( |f_{ij}(y)| |\nabla L(y)| + |\nabla f_{j}(y)| |L(y)|).
\end{equation}
The same estimate holds for the Lipshitz constant of $\bar H$. Therefore, (\ref{coverestH}) and (\ref{gradtH}) give \begin{equation} \label{complete}
\sup_{y \in {\mathbb V}} |H(y)- \bar H(y) | < \epsilon/4 + r C_3 \epsilon,
\end{equation} where $C_3>0$ is a constant uniform in $m$. Let us set $r=1/4C_3$. Then condition (\ref{LFeps}) of (\ref{mainTL}) and (\ref{complete}) show that  (\ref{LFeps2}) is satisfied. Furthermore, to complete the proof,
we compute the probability that all the events defined by (\ref{coverest}) take place by estimates analogous to (\ref{canal}), (\ref{Chern}) and (\ref{Chern2}).
\bibliography{planetary}


\end{document}